\documentclass[10pt]{article}
\usepackage{amsmath}
\usepackage{braket}
\usepackage{amssymb}
\usepackage{graphicx}
\usepackage[utf8]{inputenc}
\usepackage{color}
\usepackage{multirow}
\usepackage{tikz}
\usetikzlibrary{arrows}
\usepackage{tkz-graph}

\usepackage{float,epsfig}

\usepackage{subcaption, multirow}

\usepackage{tkz-graph}

\usepackage{amsthm}

\usepackage{color,graphicx,float,lscape,enumerate}

\usepackage[colorlinks = true]{hyperref}

\usepackage{cleveref}

\begin{document}

\newtheorem{theorem}{\bf Theorem}[section]
\newtheorem{proposition}[theorem]{\bf Proposition}
\newtheorem{definition}[theorem]{\bf Definition}
\newtheorem{corollary}[theorem]{\bf Corollary}
\newtheorem{example}[theorem]{\bf Example}
\newtheorem{exam}[theorem]{\bf Example}
\newtheorem{remark}[theorem]{\bf Remark}
\newtheorem{lemma}[theorem]{\bf Lemma}
\newtheorem{conjecture}[theorem]{\bf Conjecture}
\newcommand{\nrm}[1]{|\!|\!| {#1} |\!|\!|}
\newcommand{\quotes}[1]{``#1''}
\newcommand\numberthis{\addtocounter{equation}{1}\tag{\theequation}}

\newcommand{\ba}{\begin{array}}
\newcommand{\ea}{\end{array}}
\newcommand{\von}{\vskip 1ex}
\newcommand{\vone}{\vskip 2ex}
\newcommand{\vtwo}{\vskip 4ex}
\newcommand{\dm}[1]{ {\displaystyle{#1} } }

\newcommand{\be}{\begin{equation}}
\newcommand{\ee}{\end{equation}}
\newcommand{\beano}{\begin{eqnarray*}}
\newcommand{\eeano}{\end{eqnarray*}}
\newcommand{\inp}[2]{\langle {#1} ,\,{#2} \rangle}
\def\bmatrix#1{\left[ \begin{matrix} #1 \end{matrix} \right]}

\def \Gr{{\mathsf{Grov}}}
\def \A{{\mathbb A}}
\def \N{{\mathbb N}}
\def \R{{\mathbb R}}
\def \C{{\mathbb C}}
\def \Q{{\mathbb Q}}
\def \Z{{\mathbb Z}}
\def \S{{\mathcal S}}
\def \calP{{\mathcal P}}
\def \calG{{\mathcal G}}
\def \calPO{{\mathcal {PO}}}
\def \calX{{\mathcal X}}
\def \calY{{\mathcal Y}}
\def \calZ{{\mathcal Z}}
\def \calW{{\mathcal W}}
\def \pf{{\bf Proof: }}
\def \lam{{\lambda}}
\def\lc{\left\lceil}   
\def\rc{\right\rceil}


\title{Periodicity of lively quantum walks on cycles with generalized Grover coin}
\author{Rohit Sarma Sarkar\thanks{Department of Mathematics, IIT Kharagpur,Email: rohit15sarkar@yahoo.com}, Amrita Mandal\thanks{Department of Mathematics, IIT Kharagpur,Email: mandalamrita55@gmail.com}, Bibhas Adhikari\thanks{Corresponding author, Department of Mathematics, IIT Kharagpur,Email: bibhas@maths.iitkgp.ac.in}}
\date{}
\maketitle

\noindent{\bf Abstract:} In this paper we extend the study of three state lively quantum walks on cycles by considering the coin operator as a linear sum of permutation matrices, which is a generalization of the Grover matrix. First we provide a complete characterization of orthogonal matrices of order $3\times 3$ which are linear sum of permutation matrices. Consequently, we determine several groups of complex, real and rational orthogonal matrices. We establish that an orthogonal matrix of order $3\times 3$ is a linear sum of permutation matrices if and only if it is permutative. Finally we determine period of lively quantum walk on cycles when the coin operator belongs to the group of orthogonal (real) linear sum of permutation matrices.  \\



\section{Introduction}
Quantum walks are quantum analogues of classical random walks \cite{Aharonov2001} \cite{Ambainis2003}. Besides being a universal model for quantum computation \cite{childs2009universal} \cite{lovett2010universal} \cite{childs2013universal}, quantum walks is an extremely useful primitive to design a plethora of quantum algorithms \cite{childs2003exponential} \cite{childs2004spatial} \cite{magniez2011search} \cite{krovi2016quantum}. Similar to their classical counterpart, quantum walks can be defined both in discrete and in continuous-time \cite{Childs2002}. In this article, we focus on the former.  In quantum walks, the walker is considered as a quantum particle and the vertices of the graph determine the position space, modeled as a Hilbert space.

Let $G$ be a graph with the set of vertices $V=\{0,1, \hdots, n-1\}$. Then consider the Hilbert space, denoted by $\mathcal{H}_p$ which is spanned by the quantum states localized at the vertices of $G$. The position of the quantum walker, localized at any vertex $i$, would then be defined as $\ket{i}$.  The local transition rule  of the walk is defined by a unitary matrix, the quantum version of a classical coin, called the coin operator. The states of the coin live in an another Hilbert space, denoted by $\mathcal{H}_c,$ called the coin space which describes the direction of the particle for the walk.  Based on the state of the coin, a shift operator, denoted by $S$ defines the quantum walk that ultimately decides the position of the particle during the evolution of the walk. Thus total state space of the quantum random walk is given by the Hilbert space $\mathcal{H}_p \otimes \mathcal{H}_c$ where $\otimes$ denotes tensor product. For a brief review on quantum walks, see \cite{kempe2003quantum} \cite{venegas2012quantum} \cite{Higuchi2012} and the references therein.

The first step of a discrete quantum walk is the action of the coin operator on the existing coin state of the particle. We denote the coin operator as $C$ which is a unitary matrix. Until recently, the coin operator is used to be considered as a unitary matrix of order $2\times 2.$ In most occasions, the Hadamard matrix is the choice for a balanced unitary coin \cite{konno2017periodicity}.  Thus the quantum walk is governed by the unitary matrix $U=S(C\otimes I),$ where $I$ denotes the identity matrix. The state of the quantum walk at time $t$ is given by $\ket{\Psi_t}=U^t\ket{\Psi_0},$ where $\ket{\Psi_0}$ denotes the initial state of the walk.  The period of a walk defined by $U$ is the smallest positive integer $T$ for which $U^T=I,$ if such a $T$ does not exist then the walk is called not periodic or the period is said to be $\infty.$ Thus a quantum walk with finite period means that the walk returns to the initial state within a finite number of steps \cite{dukes2014quantum}. This property of a discrete quantum walk on cycle distinguishes itself from its classical counterpart since classical random walk on a cycle return to its starting state at irregular, unpredictable times \cite{tregenna2003controlling}.

Determining periodicity of quantum walks is one of the problems of paramount interest due to the fact that periodicity is closely related to perfect state transfer  \cite{godsil2011periodic} \cite{higuchi2017periodicity} \cite{konno2017periodicity}  \cite{saito2018periodicity} \cite{Kajiwara2019}. Indeed, periodicity is a necessary condition for perfect state transfer over symmetric graphs for discrete time walks \cite{barr2012periodicity}. Periodicity of a discrete quantum walk on cycle was first observed in \cite{travaglione2002implementing}. Periodicity of two state quantum walks on regular graphs with Fourier coin is investigated in \cite{saito2018periodicity}. Periodicity also plays an important role in experimental realization of quantum walks on a cycle \cite{saito2018periodicity} \cite{melnikov2016quantum}.    


Recently, quantum walks with high dimensional coin operator are introduced in literature \cite{brun2003quantum} \cite{inui2005one} \cite{miyazaki2007wigner} \cite{kollar2010recurrences} \cite{stefanak2014stability}. One immediate application of such coin operators is that coin degrees of freedom can offer controls over the evolution of the walk \cite{tregenna2003controlling}. Often, for three state quantum walk, the coin operator is chosen as the Grover matrix or Fourier matrix of dimension $3\times 3.$ For example, one of the well-known $3$-state quantum walks is called lively quantum walk on cycle, is introduced in \cite{Sadowski2016}. In this walk, depending on the state of the coin, the particle moves to a neighbor of the vertex or can make a jump to a vertex up to a certain distance. The distance of the jump is called the liveliness parameter of the walk. If the value of this parameter is zero then the particle does not make the jump and remains in its position, called the lazy quantum walk. It is also well known that the lively quantum walk has applications for the detection of link failures in a network. Besides, the lively walk enables to avoid trapping of the walker for any coin operator \cite{Sadowski2016}. Periodicity of lively and lazy quantum walks are derived in \cite{Kajiwara2019} when the coin operator is Grover or Fourier matrix, and it is proved that the walk has a finite period if and only if the cycle has length $3.$ 

 Here we mention that, attempts are made in literature to define new coin operators by exploiting the eigenvector properties of the Grover matrix. For instance, a family of three-state quantum walks on line is proposed in \cite{vstefavnak2012continuous} such that the coin operators of the walk posses the localization property of the Grover walk. In this case, the coin operators are specific one-parameter unitary matrices which are linear sum of the projectors corresponding to the eigenvectors of the Grover matrix.

In this work, we consider lively discrete-time quantum walks on a cycle and determine its periodicity when the underlying coin operator is a generalization of the Grover Matrix. Our motivation for this is two fold: Firstly, this would help in an improved understanding of the underlying quantum dynamics and how it differs from its classical counterpart. Secondly, from a mathematical point of view, the Grover matrix is a sum of permutation matrices. We consider as coin operator any $3\times 3$ real orthogonal matrix that is linear sum of permutation matrices, thereby generalizing the well known Grover coin operator. Moreover, characterizing unitary matrices that are linear sum of permutation matrices has been an unsolved problem. Only a necessary condition is known for a linear sum of permutation matrices to be orthogonal \cite{Kapoor1975}. In this article, we are able to completely characterize all complex, real, rational orthogonal matrices of dimension $3\times 3$, which are linear sum of permutation matrices. We establish that a linear sum of $3\times 3$ permutation matrices is orthogonal if and only if it is a permutative orthogonal matrix. In contrast to the earlier results on periodicity of lively quantum walk on cycle corresponding to Grover or Fourier coin that the period can be finite if and only if the length of the cycle is $3,$ we show that the period can be finite for cycles of any length when specific coin operators are chosen from the orthogonal group of linear sum of permutation matrices.

\section{Preliminaries}\label{sec:2}
In this section we briefly discuss some of the fundamental concepts of algebra for the completeness of the presentation of our results and that will be used in sequel. We follow the notations and terminologies given in \cite{ash2000abstract}.  A field $E$ is said to be an {\it extension} of a field $F$ if $F\subseteq E.$ We denote it by $E/F.$ Then it follows that $E$ is a vector space over the field $F,$ and if the dimension of this vector space is finite, denoted by $[E : F]$ then $E$ is called a finite field extension of $F.$ If any $f\in F[x]$ can be written as $f(x)=\lam (x-\alpha_1)\hdots (x-\alpha_k)$ for some $\alpha_1,\hdots,\alpha_k\in E, \lam\in F$ then we say that $f$ splits over $E.$ The field $E$ is called a {\it splitting field} for an $f\in F[x]$ if $f$ splits over E but not over any proper subfield of $E$ that contains $F.$ An element $\alpha\in E$ is said to be {\it algebraic} over $F$ if $\alpha$ is a root of a nonconstant polynomial $f(x)\in F[x].$ If every element of $E$ is algebraic over $F$ then $E$ is called an algebraic extension of $F.$ An algebraic extension field $E/F$ is called {\it normal} if every irreducible $f\in F[x]$ which has at least one root in $E,$ splits over $E.$

Given an element $\alpha\in E/F$ that is algebraic over $F,$ the set $\mathcal{I}$ of polynomials $f\in F[x]$ such that $f(\alpha)=0$ forms an {\it ideal} in $F[x].$ Further, since $F[x]$ is a {\it principal ideal domain}, $\mathcal{I}$ is generated by a single polynomial, say $m(x).$ If $m(x)$ is monic then it is called the minimal polynomial of $\alpha$ over $F.$  An element $\alpha\in E$ is called separable over $F$ if $\alpha$ is algebraic over $F$ and the minimal polynomial of $\alpha$ over $F$ is a separable polynomial. A field $E/F$ is called a separable extension of $F$ if every element of $E$ is separable over $F.$ An extension $E/F$ is called a Galois extension if $E/F$ is normal and separable.

Let $E/F$ be a Galois extension. Then the set of all field isomorphisms $f: E\rightarrow E$ such that $ f(x)=x, x\in F$ forms a group under the operation of composition of functions. This group is called the Galois group, denoted by $\mbox{Gal}(E/F).$ An extension $E/F$ is called cyclic extension if $E/F$ is a Galois extension and $\mbox{Gal}(E/F)$ is cyclic. 

Further, since $E$ is a vector space over $F$ when $E/F,$ we can define $F$-linear transformation $M: E\rightarrow E.$ Then norm of an element $x\in E$ is defined as $N(x)=\det(M).$ 

For example, $\Q(\sqrt{d})$ is extension of $\Q$, where $d$ is a square-free integer. Then the minimal polynomial of $\sqrt{d}$ over $\Q$ is $m(x)=x^2 - d.$ Besides, $\Q(\sqrt{d})/\Q$ is Galois and elements of the $\mbox{Gal}(\Q(\sqrt{d})/\Q)$ are $\sigma:\Q(\sqrt{d}) \rightarrow \Q(\sqrt{d})$ defined by $\sigma(a+\sqrt{d} b)=a-\sqrt{d} b,$ and the identity function. Besides, $N(a+\sqrt{d} b)=a^2-b^2d.$ 

\begin{theorem}[Hilbert's theorem 90]
If $E/F$ is a cyclic extension with $[E : F] = n$ and Galois group $G = \{1, \sigma, \hdots , \sigma_{n-1}\}$ generated by $\sigma$, and $x\in E$, then: $N(x)=1$ if and only if there exists $y\in E$ such that $x = y/\sigma(y).$ 
\end{theorem}

An extension $E/F$ is called cyclotomic if $E$ is a splitting filed for $f(x)=x^n -1\in F[x].$ The roots of $f(x)$ are called (primitive) $n$th roots of unity, denoted by $\zeta_n$. We denote a cyclotomic field by $F[\zeta_n]$ for any positive integer $n.$ 

Let $R$ be a commutative ring and $S$ be a subring of $R.$ Then an element $x\in R$ is said to be integral over $S$ if $f(x)=0$ for some monic polynomial $f\in S[x].$ If $R=\C$ and $x\in \C$ is integral over $\Z$ then $x$ is called an algebraic integer. Let $\A$ be the set of algebraic integers. Then we denote $\Z[\zeta_n]=\A\cap \Q[\zeta_n]$ where $\C/\Q.$ 

The Euler's totient function $\phi(m)$ defines the number of positive integers that are less than $m$ and relatively prime to $m.$ Any $\tau \in \mathbb{Z}[\zeta_n]$ can be written as $\sum_{i=0}^{\phi(n)-1} \zeta_n^i$ \cite{Kajiwara2019}.

\section{$3\times 3$ orthogonal matrices which are linear sum of permutations}\label{sec:3}

In this section we provide a complete characterization of $3\times 3$ orthogonal matrices which are linear sum of permutation matrices. First we characterize permutative  orthogonal matrices of order $3\times 3$, and then we show that linear sum of permutation matrices are orthogonal if and only if they are orthogonal permutative matrices. Besides, we characterize all rational permutative orthogonal matrices of order $3\times 3.$ 

\subsection{Complex and real matrix groups of linear sum of permutation matrices}

Recall that a matrix is permutative if the rows of the matrix are permutations of each other \cite{paparella2015realizing}. Thus any $3\times 3$ permutative matrix is of the form \begin{equation} A({\bf x};P,Q)=\bmatrix{{\bf x}\\ {\bf x}P\\{\bf x}Q}\end{equation} where ${\bf x}=\bmatrix{x & y & z }$ is a row vector in $\R^3$ and $P, Q\in\calP,$ which is the group of permutation matrices of order $3\times 3.$ The following theorem provides a characterization of all permutative orthogonal matrices of order $3\times 3$ which we denote by $\calPO.$ Obviously $\calP\subset\calPO.$ The elements of $\calP$ are matrices corresponding to the permutations in the symmetric group $S_3.$ If $\pi\in S_3$ then the permutation matrix corresponding to $\pi$ is defined as $P=[p_{ij}]$ where $p_{ij}=1$ if and only if $\pi(i)=j.$ Thus $\calP$ can be described as \begin{equation}\label{des:perm} \calP = \{P_1=I=\mbox{id}, P_2=(123), P_3=(132), P_4=(23), P_5=(12), P_6=(13)\}\end{equation} where $\mbox{id}$ denotes the identity element of the symmetric group. 

\begin{theorem} \label{per_ortho}
A matrix $A({\bf x}; P,Q)\in \calPO$ if and only if it is either $x P_1 + y P_2 + z P_3$ or $x P_4 + y P_5 + z P_6$ where $(x,y,z)$ belongs to either of the affine varieties
\begin{equation}\label{eqn:av}{\bf S_+}= \left\{
  \begin{array}{ll}
    x+y+z=1  \\
    x^2+y^2+z^2=1
  \end{array}
\right. \,\, \mbox{and} \,\, {\bf S_-}=\left\{
  \begin{array}{ll}
    x+y+z=-1  \\
    x^2+y^2+z^2=1
  \end{array}
\right.\end{equation} 
and $P_i, i=1,\hdots,6$ are described in (\ref{des:perm}).
\end{theorem}

\noindent{\bf Proof:} The `if' part is obvious and easy to check. To prove the `only if' part consider the following cases. First assume that $A({\bf x}; P,Q)\in \calPO$ and the symbolic matrix $A({\bf x}; P,Q)$ has no repetition of entries in any of the columns. Then $A({\bf x}; P,Q)$ can be either of the following forms. Besides, since rows and columns are orthogonal, $P\neq Q$ and $P,Q$ are not the identity matrix. Thus
 \beano A({\bf x};P_2,P_3) &=& \bmatrix{
x & y & z\\
z & x & y\\
y & z & x\\
}=x P_1+ yP_2+ zP_3, \, \mbox{and} \\ A({\bf x};P_3,P_2) &=& \bmatrix{x & y&z\\y&z&x\\z&x&y}=xP_4+yP_5+zP_6. \eeano

Further, due to the orthonormality condition of the rows, it follows that $xy+yz+xz=0$ and $x^2+y^2+z^2=1$ which further imply that $x+y+z\in\{\pm 1\}.$ Thus the desired result follows.


Next, consider the symbolic orthogonal permutative matrices $A({\bf x}; P,Q),$ in which, one entry is repeated in at least one column. For any such matrix, the following system of polynomial equations hold due to the orthonormality condition of the rows. 
\beano
x_1^2+x_2^2+x_3^2 &=&1 \\
2 x_i^2 + x_j^2 &=& 1 \\
2 x_k^2+ x_j^2 &=& 1,
\eeano
where $x_i,x_j,x_k \in \{x,y,z\}$ and $x_i\neq x_j\neq x_k.$ Then solving these equations, it follows that $x_i=\pm x_k.$ Now, if $x_i=x_k$ then either $x_j\in\{\pm 1\}, x_i=x_k=0,$ or $x_j=\pm 1/3, x_i=x_k=\mp 2/3.$ Finally, if $x_i=-x_k$ then $x_j\in\{\pm 1\}$ and $x_i=x_k=0$ or $x_j=\pm 1/3, x_i=x_k=\mp 2/3.$ Then it follows that $A({\bf x}; P,Q)$ has the desired form. This completes the proof.  $\hfill{\square}$

Now we prove the main theorem of this section that characterizes all orthogonal matrices which are linear combination of permutation matrices.

\begin{theorem} \label{linear_combi_classi}
A linear combination of  $3\times 3$ permutation matrices is orthogonal  if and only if it is a permutative orthogonal matrix.
 \end{theorem}
\noindent {\bf Proof:} Let $M=\sum_{i=1}^6 \alpha_i P_i$ be a linear combination of permutation matrices. Then $M$ can be written as a sum of symmetric permutative matrix $A$ and a permutative matrix $B,$ where  $$A=\bmatrix{\alpha_4 & \alpha_5 & \alpha_6\\ \alpha_5 & \alpha_6 & \alpha_4\\ \alpha_6 & \alpha_4 & \alpha_5}, \,\, B=\bmatrix{\alpha_1 & \alpha_2 & \alpha_3\\ \alpha_3 & \alpha_1 & \alpha_2\\ \alpha_2 & \alpha_3 & \alpha_1}.$$ Then it follows that  $A^{T}B=B^{T}A,AB^{T}=BA^{T}$ and $B^{T}B=BB^{T}.$

Suppose that $M$ is orthogonal. Then $MM^T=M^TM=I$ implies  \begin{eqnarray} \label{eq1}
 AA^{T}+AB^{T}+BA^{T}+BB^{T}=A^{T}A+A^{T}B+B^{T}A+B^{T}B=I.
 \end{eqnarray} Which further implies that $A^{T}B-AB^{T}=0.$ Consequently,  $$A(B-B^{T})=(\alpha_2-\alpha_3)\bmatrix{\alpha_6-\alpha_5&\alpha_4-\alpha_6&\alpha_5-\alpha_4\\\alpha_4-\alpha_6&\alpha_5-\alpha_4&\alpha_6-\alpha_5\\\alpha_5-\alpha_4&\alpha_6-\alpha_5&\alpha_4-\alpha_6}=0.$$ Then two cases arise: either $\alpha_2=\alpha_3$ or $\alpha_4=\alpha_5=\alpha_6.$

If $\alpha_2=\alpha_3$ then $$M=\bmatrix{\alpha_4+\alpha_1 & \alpha_5+\alpha_2 & \alpha_6+\alpha_2\\ \alpha_5+\alpha_2 & \alpha_6+\alpha_1 & \alpha_4+\alpha_2\\ \alpha_6+\alpha_2 & \alpha_4+\alpha_2 & \alpha_5+\alpha_1}.$$ Since each row has unit $2$-norm, $\alpha_i, i=1,\hdots,6$ must satisfy the following system of polynomial equations.
 \begin{eqnarray}
 \label{eq2} &(\alpha_4+\alpha_1)^2+( \alpha_5+\alpha_2)^2+(\alpha_6+\alpha_2)^2=1 \\
 \label{eq3} &(\alpha_5+\alpha_2)^2+(\alpha_6+\alpha_1)^2+(\alpha_4+\alpha_2)^2=1 \\
 \label{eq4} &(\alpha_6+\alpha_2)^2+(\alpha_4+\alpha_2)^2+(\alpha_5+\alpha_1)^2=1.
 \end{eqnarray}
Then (\ref{eq2})-(\ref{eq3}), (\ref{eq2})-(\ref{eq4}) and (\ref{eq3})-(\ref{eq4}) imply $$(\alpha_4-\alpha_6)(\alpha_1-\alpha_2)=0, (\alpha_4-\alpha_5)(\alpha_1-\alpha_2)=0, \,\mbox{and}\, (\alpha_5-\alpha_6)(\alpha_1-\alpha_2)=0$$ respectively. Thus if $\alpha_1\neq\alpha_2$ then $\alpha_4=\alpha_5=\alpha_6.$ This yields, $$M=\bmatrix{\alpha_4+\alpha_1 & \alpha_4+\alpha_2 & \alpha_4+\alpha_2\\ \alpha_4+\alpha_2 & \alpha_4+\alpha_1 & \alpha_4+\alpha_2\\ \alpha_4+\alpha_2 & \alpha_4+\alpha_2 & \alpha_4+\alpha_1}$$ which is a permutative matrix. Similarly, if $\alpha_1=\alpha_2$ then $M$ is a permutative matrix.

Finally, let $\alpha_2\neq \alpha_3.$ Then $\alpha_4=\alpha_5=\alpha_6$ which implies that $M=\alpha_4 J + B.$ Thus $M$ is a permutative matrix. 

Now consider the converse part. Let $M$ be a $3\times 3$ permutative orthogonal matrix. Then by Theorem \ref{per_ortho} it follows that $M$ is a linear combination of permutation matrices. This completes the proof. $\hfill{\square}$


Setting $z=1-x-y$ and eliminating $z$ from the equation $x^2+y^2+z^2=1,$ the $(x,y,z)$ which satisfy the system of polynomials ${\bf S_+}$ are given by $(x,y)$ which lie on the ellipse given by \begin{equation}\label{eqn:el1} x^2+y^2-x-y+xy =0.\end{equation} Similarly, the points $(x,y,z)$ which satisfy the system ${\bf S_-}$ are given by $z=-1-x-y$ and $(x,y)$ which lie on the ellipse  \begin{equation}\label{eqn:el2} x^2+y^2+x+y+xy =0.\end{equation} Hence we have the following corollary.
 
\begin{corollary}\label{cor:char}
An orthogonal matrix $M$ is a linear combination of permutation matrices of order $3\times 3$ if and only if $$M=xP_1 + yP_2 + zP_3 \,\, \mbox{or} \,\, M=xP_4 + yP_5 + zP_6$$ where $(x,y,z)$ are given by  \begin{equation}\label{eqn:av2} \left\{
  \begin{array}{ll}
    z=1-x-y  \\
  x^2+y^2-x-y+xy =0
  \end{array}
\right. \,\, \mbox{or} \,\, \left\{
  \begin{array}{ll}
    z=-1-x-y  \\
    x^2+y^2+x+y+xy =0.
  \end{array}
\right.\end{equation}
\end{corollary}

Moreover, it can further be noted that the ellipses given by equations (\ref{eqn:el1}) and (\ref{eqn:el2}) are inherently different. Observe that   \begin{eqnarray}
&& x^2+y^2-x-y+xy= X^TQ_+X=0 \,\, \mbox{and} \label{eqn1:elp}\\
&& x^2+y^2+x+y+xy = X^TQ_-X = 0,\label{eqn2:elp}
\end{eqnarray}
where $$Q_+=\bmatrix{1 & 1/2 & -1/2\\ 1/2 & 1 & -1/2\\ -1/2 & -1/2 & 0}, Q_-= \bmatrix{1 & 1/2 & 1/2\\ 1/2 & 1 & 1/2\\ 1/2 & 1/2 & 0}, X=\bmatrix{x \\ y \\1}.$$ In addition, since $-\,\mbox{coefficient of}\,\, y^2 \times \det(Q_+)>0,$ the equation (\ref{eqn1:elp}) represents a real ellipse, whereas (\ref{eqn2:elp}) represents an imaginary ellipse \cite{lawrence2013catalog}.

It follows from the proof of Theorem \ref{per_ortho} that the  permutative orthogonal matrices which have repeated entries in their columns are given by permutation matrices and matrices of the form $$\frac{2}{3} J - P \,\, \mbox{and} \,\, -\frac{2}{3} J + P$$ where $J$ is the all-one matrix and $P\in\calP.$ In particular, setting $P=I,$ the identity matrix, we obtain the Grover matrix given by $$\frac{2}{3} J - I=\bmatrix{-\frac{1}{3}& \frac{2}{3} & \frac{2}{3}\\\frac{2}{3} &-\frac{1}{3}& \frac{2}{3}\\\frac{2}{3} & \frac{2}{3}&-\frac{1}{3}}.$$

Now we have the following definition.

\begin{definition}(Grover-type matrix)\label{def:gtype}
A permutative orthogonal matrix of the form $\frac{2}{3}J-P, P\in\calP$ is called a Grover-type matrix. The set of all Grover-type matrices is denoted by $$\mathsf{Grov}= \left\{\frac{2}{3}J-P : P\in\calP\right\}.$$
\end{definition} 

It may also be noted that matrices in $\mathsf{Grov}$ can be obtained by permuting the rows of the Grover matrix. Besides,  permutation matrices, Grover-type matrices, and matrices of the form $-\frac{2}{3}J+P, P\in\calP$ belong to the classes of matrices given by Corollary \ref{cor:char}. Hence we consider the sets of all parametric orthogonal matrices which are linear combinations of permutation matrices, and investigate the algebraic structures of these sets.  

Set \begin{eqnarray}
\calX &=& \left\{ \bmatrix{
x & y &1-x-y\\
1-x-y & x & y\\
y &1-x-y & x
} : x^2+y^2-x-y+xy=0 \right\}\label{eqn:x}\\
\calY &=&  \left\{ \bmatrix{
x & y &-1-x-y\\
-1-x-y & x & y\\
y &-1-x-y & x
} : x^2+y^2+x+y+xy=0 \right\} \label{eqn:y}\\
\calZ &=&  \left\{ \bmatrix{
x & y &1-x-y\\
y & 1-x-y & x\\
1-x-y & x & y
} : x^2+y^2-x-y+xy=0 \right\} \label{eqn:z}\\
\calW &=& \left\{ \bmatrix{
x & y &-1-x-y\\
y & -1-x-y & x\\
-1-x-y & x & y
} : x^2+y^2+x+y+xy=0 \right\}. \label{eqn:w}
\end{eqnarray}

It may be noted that $\calZ=\{P_4A : A\in\calX\}$ and $\calW = \{P_4A : A\in \calY\}.$ Then we have the following proposition which describes group structures of the above sets.
\begin{proposition}\label{Thm:crgroups}
Let $\calX,\calY,\calZ$ and $\calW$ be defined as above. Then the following are true. \begin{itemize}
\item[(a)] $\det(M)=1$ if $M\in \calX, \calZ;$ and $\det(M)=-1$ if $M\in\calY, \calW.$
\item[(b)] $\calX, \calX\cup\calY, \calX\cup\calZ, \calX\cup\calW, \calX\cup\calY\cup\calZ\cup\calW=\mathcal{PO}$ are complex orthogonal matrix groups with respect to matrix multiplication, and $\calX$ is a normal subgroup of all these groups.
\item[(c)]  $\calX$ is a real orthogonal matrix group, denoted by $\calX_R$ if and only if $-\dfrac{1}{3}\leq x \leq 1.$
\item[(d)] $\calX_R\cup\calY_R$ is a real orthogonal matrix group where $\calY_R=\{A\in\calY : -1\leq x\leq \dfrac{1}{3}\}.$
\item[(e)] $\calX_R\cup \calZ_R$  is a real orthogonal matrix group where $\calZ_R=\{A\in\calZ :-\dfrac{1}{3}\leq x \leq 1\}.$
\item[(f)] $\calX_R\cup \calW_R$  is a real orthogonal matrix group where $\calW_R=\{A\in\calW : -1\leq x\leq \dfrac{1}{3}\}.$
\item[(g)] $\cal{PO}_R=\calX_R\cup \calY_R\cup\calZ_R\cup\calW_R$  is a real orthogonal matrix group. 
\end{itemize}
\end{proposition}
\noindent{\bf Proof:} Proof of $(a)$ is computational and easy to verify. Consider $(b).$ First we prove that $\calX\cup\calY\cup\calZ\cup\calW$ is a complex orthogonal matrix group. Clearly $I\in\calX\cup\calY\cup\calZ\cup\calW.$ Since any $A\in\calX\cup\calY\cup\calZ\cup\calW$ is orthogonal, $A^{-1}=A^T.$ If $A\in\calX$ then $A^T\in \calX$ follows from exchanging the role of $1-x-y$ and $y.$ Similarly, $A^T\in\calY$ if $A\in\calY.$ Since $\calZ$ and $\calW$ contain complex symmetric matrices, obviously $A^T\in\calZ$ if $A\in\calZ$, and $A^T\in\calW$ if $A\in\calW.$ Hence, $\calX\cup\calY\cup\calZ\cup\calW$ is closed under inverses. The closure property follows by considering the following cases. 

Let $A=x_1P_1+y_1P_2+z_1P_3, B=x_2P_1+y_2P_2+z_2P_3\in\calX\cup\calY$ where $(x_i,y_i)$ satisfies one of the equations (\ref{eqn:el1}) and (\ref{eqn:el2}), and $z_i=1-x_i-y_i, i\in\{1,2\}$. Note that $P_2^2=P_3, P_3^2=P_2, P_2P_3=I=P_3P_2.$ Then 
$AB = x_3P_1 + y_3P_2 + z_3P_3$ where $$x_3=x_1x_2+y_1z_2+z_1y_2, \,\, y_3=x_1y_2+x_2y_1+z_1z_2, \,\, z_3=x_1z_2+y_1y_2+z_1x_2.$$ Then $z_3=1-x_3-y_3$ if $z_i=1-x_i-y_i$ or $z_i=-1-x_i-y_i,$ $i=1,2,$ and $z_3=-1-x_3-y_3$ if $z_i=1-x_i-y_i$ and $z_j=-1-x_j-y_j,$ $i\neq j, i,j\in\{1,2\}.$ Hence $x_3+y_3+z_3\in\{\pm 1\}.$ Further, \beano x_3^2+y_3^2+z_3^2 &=& \left(\bmatrix{x_1 \\ y_1 \\ z_1}^T\bmatrix{x_2\\z_2\\y_2}\right)^2 +  \left(\bmatrix{x_1 \\ y_1 \\ z_1}^T\bmatrix{y_2\\x_2\\z_2}\right)^2 +  \left(\bmatrix{x_1 \\ y_1 \\ z_1}^T\bmatrix{z_2\\y_2\\x_2}\right)^2 \\ &=& \bmatrix{x_1\\y_1\\z_1}^T \left(\bmatrix{x_2\\z_2\\y_2}\bmatrix{x_2\\z_2\\y_2}^T\right)\bmatrix{x_1\\y_1\\z_1} +  \bmatrix{x_1\\y_1\\z_1}^T \left(\bmatrix{y_2\\x_2\\z_2}\bmatrix{y_2\\x_2\\z_2}^T\right)\bmatrix{x_1\\y_1\\z_1} \\ && +  \bmatrix{x_1\\y_1\\z_1}^T \left(\bmatrix{z_2\\y_2\\x_2}\bmatrix{z_2\\y_2\\x_2}^T\right)\bmatrix{x_1\\y_1\\z_1}  \\ &=&  \bmatrix{x_1\\y_1\\z_1}^T  \bmatrix{a & b &b\\ b &a & b\\ b & b & a}\bmatrix{x_1\\y_1\\z_1},\eeano where $a=x_2^2+y_2^2+z_2^2=1$ and   $b=x_2z_2+x_2y_2+y_2z_2=0.$ Now, $x_2+y_2+z_2\in\{\pm 1\}$ since $B\in\calX\cup\calY.$ This yields $x_3^2+y_3^2+z_3^2 = x_1^2+y_1^2+z_1^2=1.$ Thus $AB\in \calX\cup\calY.$

Now let $A = x_1P_1 + y_1P_2 + z_1P_3 \in \mathcal{X};B = x_2P_4 + y_2P_5 + z_2P_6 \in  \mathcal{Z} \cup \mathcal{W}$ where $(x_1, y_1)$ satisfies the equation (\ref{eqn:el1}), and $(x_2,y_2)$ lies on one of the ellipses given by equations (\ref{eqn:el1}) and (\ref{eqn:el2}). Thus $z_1 = 1 -x_1 -y_1$ and $x_2+y_2+z_2\in\{1,-1\}$. Note that $P_2P_4=P_5,P_2P_5=P_6,P_2P_6=P_4,P_3P_4=P_6,P_3P_5=P_4,P_3P_6=P_5.$ Then $AB = x_3P_4 + y_3P_5 + z_3P_6,$ where $$x_3=x_1x_2+y_1z_2+y_2z_1,y_3=x_1y_2+x_2y_1+z_1z_2,z_3=x_1z_2+y_1y_2+x_2z_1.$$
Hence, $$
  x_3+y_3+z_3=(x_1+y_1+z_1)(x_2+y_2+z_2)=  \left\{
  \begin{array}{ll}
   1 \,\mbox{if} \, ~B\in\mathcal{Z} \\
  -1  \,\mbox{if} \, ~B\in\mathcal{W}.
  \end{array}\right.
$$ Next, as above, $x_3^2+y_3^2+z_3^2=x_1^2+y_1^2+z_1^2=1.$ Thus $AB \in \mathcal{X}\cup \mathcal{Z}$ if $B \in \mathcal{Z}$ and $AB \in \mathcal{X} \cup \mathcal{W}$ if $B\in \mathcal{W}.$


Next let $A = x_1P_4 + y_1P_5 + z_1P_6$ and $B = x_2P_4 + y_2P_5 + z_2P_6$ such that either both $A, B$ belong to $ \mathcal{Z}$ or both belong to $\mathcal{W}.$ Thus $(x_i, y_i)$ satisfies either equation (\ref{eqn:el1}) or (\ref{eqn:el2}), and $ x_i +y_i+z_i=1$ or $-1, i=1,2$. Now,
$P_4^2=P_5^2=P_6^2=P_1, P_4P_5=P_3,P_4P_6=P_2,P_5P_4=P_2,P_5P_6=P_3,P_6P_4=P_3,P_6P_5=P_2.$ Then $AB = x_3P_1 + y_3P_2 + z_3P_3,$ where
$$x_3=x_1x_2+y_1y_2+z_1z_2,y_3=x_2y_1+x_1z_2+y_2z_1,z_3=x_1y_2+x_2z_1+y_1z_2.$$
Hence, $x_3+y_3+z_3=(x_1+y_1+z_1)(x_2+y_2+z_2)=1,$ and $x_3^2+y_3^2+z_3^2=x_1^2+y_1^2+z_1^2=1$ as above, and hence $AB \in \mathcal{X}.$ 


Therefore both $\mathcal{X}\cup \mathcal{Z}$ and $\mathcal{X}\cup \mathcal{W}$ are closed under matrix multiplication. Similarly, it can be proved that if $A,B \in \mathcal{Y}\cup\mathcal{Z}$ or $A,B \in \mathcal{Y}\cup\mathcal{W}$ then $AB\in \mathcal{X} \cup \mathcal{Y} \cup \mathcal{Z}\cup\mathcal{W}.$ Besides, it can similarly be shown that  $AB\in \mathcal{X} \cup \mathcal{Y}$ if $A,B \in \mathcal{Z}\cup\mathcal{W}.$ Thus the desired results follow from the fact that matrix multiplication is associative.

Now consider $(c).$ It is enough to determine for what real values of $x$ the value of $y$ is real such that $(x,y)\in\R^2$ which satisfy equation (\ref{eqn:el1}). Consider the equation (\ref{eqn:el1}) as a polynomial equation in which $y$ is the indeterminate. Then $$y = \dfrac{1}{2}\left[ (1-x) \pm\sqrt{(1-x)^2 - 4(x^2-x)} \right].$$ Then it follows that $y$ is real if and only if $x$ lies in the closed interval $[-1/3, \,\, 1].$ Indeed, $y\in [-1/3, \,\, 1].$ The proof of $(d)-(g)$ follow similarly. This completes the proof. 
$\hfill{\square}$

From the construction of real orthogonal matrix groups discussed in the above theorem it follows that the matrices in $\calX_R$ ultimately depend on the parameter $x$ which acts as a free parameter in a bounded interval. Then $y,$ and hence $z$ can be written as functions of $x.$ Based on this observation we parametrize the values of $x$ in the domain of the interval. Thus we have the following corollary.

\begin{corollary}
The following sets are one-parameter orthogonal matrix groups.\begin{itemize} 
\item[(a)] $$\calX_\theta=\left\{\bmatrix{
\frac{2\cos{\theta}+1}{3} & \frac{(1-\cos{\theta})}{3}+ \frac{1}{\sqrt{3}}\sin{\theta} & \frac{(1-\cos{\theta})}{3}- \frac{1}{\sqrt{3}}\sin{\theta}\\
\frac{(1-\cos{\theta})}{3} - \frac{1}{\sqrt{3}}\sin{\theta} & \frac{2\cos{\theta}+1}{3} & \frac{(1-\cos{\theta})}{3} + \frac{1}{\sqrt{3}} \sin{\theta}\\
\frac{(1-\cos{\theta})}{3} + \frac{1}{\sqrt{3}} \sin{\theta} & \frac{(1-\cos{\theta})}{3} - \frac{1}{\sqrt{3}} \sin{\theta} & \frac{2\cos{\theta}+1}{3}
} : 0\leq \theta < 2\pi \right\}.$$ 
\item[(b)] $\calX_\theta\cup \calY_\theta$ where $$\calY_\theta = \left\{ \bmatrix{
\frac{(2\cos{\theta}-1)}{3} & -\frac{(1+\cos{\theta})}{3}+ \frac{1}{\sqrt{3}}\sin{\theta} & -\frac{(1+\cos{\theta})}{3}- \frac{1}{\sqrt{3}}\sin{\theta}\\
-\frac{(1+\cos{\theta})}{3} - \frac{1}{\sqrt{3}}\sin{\theta} & \frac{(2\cos{\theta}-1)}{3} & -\frac{(1+\cos{\theta})}{3} + \frac{1}{\sqrt{3}}\sin{\theta}\\
-\frac{(1+\cos{\theta})}{3} + \frac{1}{\sqrt{3}} \sin{\theta} & -\frac{(1+\cos{\theta})}{3} - \frac{1}{\sqrt{3}} \sin{\theta} & \frac{(2\cos{\theta}-1)}{3}} : 0\leq \theta < 2\pi\right\}.$$
\item[(c)] $\calX_\theta\cup\calZ_\theta$ where $\calZ_\theta=\{P_4A : A\in\calX_\theta\},$ that is, matrices in $\calZ_\theta$ are matrices in $\calX_\theta$ with second and third row exchanged. 

\item[(d)] $\calX_\theta\cup\calW_\theta$ where $\calW_\theta=\{P_4A : A\in\calY_\theta\},$ that is, matrices in $\calW_\theta$ are matrices in $\calY_\theta$ with second and third row exchanged.

\item[(e)] $\calPO_\theta := \calX_\theta\cup\calY_\theta\cup\calZ_\theta\cup\calW_\theta.$ 
\end{itemize}
\end{corollary}
\noindent{\bf Proof:} The one parameter representations of $\calX_R, \calY_R, \calZ_R$ and $\calW_R$ are obtained as follows. $\calX_\theta$ follows from the construction of $\calX_R$ by setting $x=\frac{1+2\cos\theta}{3},$ and hence $y= \frac{(1-\cos{\theta})}{3}+ \frac{1}{\sqrt{3}}\sin{\theta}.$ $\calY_R$ is the set all matrices in $\calY$ in which the parameter $x$ lies in the interval $[-1, \,\, 1/3].$ Then setting $x=\frac{2\cos\theta-1}{3}$, $\calY_\theta =\calY_R.$
Similarly, the expressions of $\calZ_\theta, \calW_\theta$ can be obtained. The rest follows from Proposition \ref{Thm:crgroups}. \hfill{$\square$}

\subsection{Rational matrix group of linear sum of permutation matrices}
In this section we determine rational orthogonal matrices which are linear sum of permutation matrices. Thus we need to determine values of $(x, y)\in\Q\times \Q$ which lie on the ellipses given by equations (\ref{eqn:el1}) and (\ref{eqn:el2}). 

First we determine the set of rational matrices, denoted by $\calX_Q \subset\calX_R.$ An equation of the form $x^2-dy^2=k,$ where $k$ is an integer and $d$ is a nonsquare positive integer, is well-known as Pell's equation.



Recall that \begin{equation}\label{eqn:xr}\calX_R=\left\{ \bmatrix{
x & y &1-x-y\\
1-x-y & x & y\\
y &1-x-y & x
} : x^2+y^2-x-y+xy=0, -\dfrac{1}{3}\leq x\leq 1 \right\}.\end{equation} Now treating $x^2+y^2-x-y+xy=0$ as a polynomial in indeterminate $y,$ we obtain $$y=\dfrac{(1-x)\pm\sqrt{(1-x)(3x+1)}}{2}.$$ Then $y\in\Q$ if and only if $(1-x)(3x+1)$ is zero or perfect square of a nonzero rational number. It is zero if $x\in\{1,-1/3\}.$ If $x=1$ then $y=0$ and hence the corresponding orthogonal matrix in $\calX_Q$ is the identity matrix, otherwise if $x=-1/3$ then $y=2/3$ which gives the Grover matrix as an element of $\calX_Q.$ If it is nonzero then $$(1-x)(3x+1)=\dfrac{p^2}{q^2}$$ for some nonzero rational number $p/q.$ This gives $x=\dfrac{q\pm\sqrt{4q^2-3p^2}}{3q}.$ 

Now we need to find values of $p, q$ for which $x\in\Q.$ Note that $4q^2-3p^2\neq 0,$ since otherwise $p/q=2/\sqrt{3}$ which is not rational. Thus $4q^2-3p^2$ has to be square of a nonzero integer, say $m.$ Hence $4q^2-3p^2=m^2$ which is a Pell's equation, which finally taks the form $$X^2-3Y^2=1$$ where $X=2q/m$ and $Y=p/m.$ Now we apply Hilbert's Theorem 90 (see Section \ref{sec:2}).

Set $E=\Q(\sqrt{3})$ and $F=\Q.$ Besides $N(x)=X^2-3Y^2=1$ where $x=X+\sqrt{3}Y\in\Q(\sqrt{3})$ as given above, and $\sigma(X+\sqrt{3} Y)=X-\sqrt{3} Y$ where $\sigma: \Q(\sqrt{3})\rightarrow\Q(\sqrt{3})$ is the generator of $\mbox{Gal}(\Q(\sqrt{3})/\Q).$ Then due to Hilbert's Theorem 90 there must exist $y=a+\sqrt{3} b\in\Q(\sqrt{3})$ such that $$X+\sqrt{3} Y =\frac{a+\sqrt{3} b}{\sigma(a+\sqrt{3} b)}=\frac{a+\sqrt{3} b}{a-\sqrt{3} b}=\frac{a^2+3b^2}{a^2-3b^2} + \sqrt{3}\frac{-2ab}{a^2-3b^2}.$$ Thus $$X=\frac{r^2+3}{r^2-3} \,\, \mbox{and} \,\, Y=\frac{-2r}{r^2-3}$$ for some $r=a/b\in\Q.$ Consequently we obtain $$q=\frac{m(r^2+3)}{2(r^2-3)} \,\, \mbox{and} \,\, p=\frac{-2mr}{r^2-3}$$ which satisfy the equation $4q^2-3p^2=m^2,$ for some nonzero integer $m.$ 

Then we have the following proposition.
\begin{proposition}
The set $\calX_Q, \calX_Q\cup\calY_Q, \calX_Q\cup\calZ_Q, \calX\cup\calW_Q,\calPO_Q=\calX_Q\cup\calY_Q\cup\calZ_Q\cup\calW_Q$ are matrix groups of orthogonal rational matrices, where \beano \calX_Q &=&\left\{  A\in\calX_R : x= \dfrac{1}{3}\pm \dfrac{2(r^2-3)}{3(r^2+3)}, \, y=\dfrac{1}{3}\mp\dfrac{r^2-3\pm 6r}{3(r^2+3)}, r\in\Q \right\}, \\ \calY_Q &=& \left\{ A\in \calY_R : x= -\frac{1}{3} \pm \frac{2(r^2-3)}{3(r^2+3)}, \,y=-\frac{1}{3} \mp \frac{r^2-3\pm 6r}{3(r^2+3)}, r\in \mathbb{Q}\right\} \\
\calZ_Q &=& \{P_4A : A\in\calX_Q\} \\
\calW_Q &=& \{P_4A : A\in\calY_Q\}.\eeano 

\end{proposition} 
\noindent{\bf Proof:} From the above discussion it follows that  \begin{eqnarray}\calX_Q &=&\left\{  \bmatrix{
x & y &1-x-y\\
1-x-y & x & y\\
y &1-x-y & x
} : x= \frac{q\pm m}{3q}, y=\frac{(1-x)\pm \frac{p}{q}}{2}\right. \nonumber\\ && \left. p=\frac{-2mr}{r^2-3}, q=\frac{m(r^2+3)}{2(r^2-3)}, r\in\Q, m\in\Z \right\}.\nonumber\end{eqnarray} Then the desired result for $\calX_Q$ can be obtained by a straightforward calculation. For $\calY_Q,$ note that the parameters $x,y$ satisfy $x^2+y^2+xy+x+y=0$. This implies $(1+x)(1-3x)=\frac{p^2}{q^2}$ and using above arguments it follows that $x= \frac{-q\pm m}{3q}$ and $y=\frac{-(1+x)\pm \frac{p}{q}}{2}.$ Thus the desired results follows for other cases using Proposition \ref{Thm:crgroups}.  \hfill{$\square$}


\section{Periodicity of lively quantum walks on cycles}

In this section we generalize the lively quantum walk on cycles by considering the coin operator as an orthogonal matrix which is a linear sum of permutation matrices. The lively quantum walk on cycles is defined as follows. The state space of the walk is $\C^3\otimes\C^n$ where $n$ is the number of nodes in the cycle $C_n,$ and $\otimes$ denotes the tensor product as usual. The evolution operator is given by \begin{equation}\label{evlop} U_{(n,a)}=S^{(n,a)}(C\otimes I), \end{equation} where $C$ is the coin operator, $S^{(n,a)}$ is the shift operator, and $a\leq \lfloor\frac{n}{2}\rfloor$ is called the liveliness parameter \cite{Sadowski2016}. If $a=0$ the corresponding walk is called the lazy walk. The shift operator is defined by \begin{equation}\label{shftop}S^{(n,a)}=\sum_{x=0}^{n-1} S_x^{(n,a)}\end{equation} where \begin{eqnarray} S_x^{(n,a)} &=& \ket{0}\bra{0}\otimes \ket{x-1(\mbox{mod}\, n)}\bra{x} \nonumber \\ && + \ket{1}\bra{1}\otimes \ket{x+1(\mbox{mod}\, n)}\bra{x} \nonumber \\ && + \ket{2}\bra{2}\otimes \ket{x+a(\mbox{mod}\, n)}\bra{x}. \end{eqnarray} Thus $S_x^{(n,a)}$ acts as shifting the position of the walking particle from the position at the vertex $x$ of to one of the positions $x-1,$ $x-1$ and $x+a$ depending on the coin state. The position state is described by $\ket{x}, x\in\{0,1,\hdots,n-1\}$ which form a set of orthonormal standard basis vectors of $\C^n.$ Obviously, the vertex set of $C_n$ is given by $\{0, 1, \hdots, n-1\}.$

The periodicity of this walk is investigated in \cite{Sadowski2016} where the coin operator $C$ is the Grover matrix $G=\frac{2}{3}J-I\in\Gr.$ Recently, periodicity of three state lazy quantum walk on cycles is considered in \cite{Kajiwara2019}. In this section we investigate the periodicity of the lively quantum walk on cycle $C_n$ when $C\in \calPO_R$ and the pair $(n,a)$ satisfies the condition $\frac{n}{\mbox{gcd}(n,a)}$ divides $l,$ for some $l\in\{0,1,\hdots,n-1\}.$ We follow a similar approach to derive the periodicity of lively quantum walks where the coin operator is an orthogonal matrix which is a linear sum of permutation matrices, as described in Section \ref{sec:3}. 
   
If $\Psi_0\in\C^3\otimes\C^n$ denote the initial state of the lively quantum walk then the discrete time evolution of the walk is given by \begin{equation}\label{def:qe} \Psi(t) = U_{(n,a)}^t\Psi_0 \end{equation} where $t\in\N$ denotes the time, and $U_{(n,a)}^0=\Psi(0)=\Psi_0.$ The walk is said to be periodic if $U_{(n,a)}^t=I$ for some $t\in\N.$ The smallest $t=T$ for which the walk is periodic, is called the period of the walk. If no such $t$ exists then the walk is called not periodic. One of the fundamental problems in the area of quantum walks is to decide if a walk is periodic and then find the period. The periodicity of a walk can be investigated through the eigenvalues of the corresponding evolution operator of the walk. It can be easily shown that $U_{(n,a)}^t=I$ for some $t$ if and only if $\lam^t=1$ for any eigenvalue $\lam$ of $U_{(n,a)}$ \cite{Kajiwara2019}.

In this paper we consider the lively quantum walk defined by the evolution operator $U_{(n,a)}$ given by the equation (\ref{evlop}), where the coin operator $C\in \calPO_R$ and the shift operator is given by equation (\ref{shftop}). The following proposition shows that eigenvalues of $U_{(n,a)}$ can be accessed from the eigenvalues of a unitary matrix $U_k=D_kC\in\C^{3\times 3}$ where $D_k$ is a unitary diagonal matrix. 

\begin{proposition} \label{l1}
Let $C\in \calPO_R.$ Then the lively quantum walk operator $U_{(n,a)}=S^{(n,a)}(C\otimes I_n)$ has eigenvalues $\lambda_{k,j}$ with a corresponding eigenvector $\ket{\psi_{k,j}}=\ket{\nu_{k,j}}\otimes \ket{\phi_k}$ where $\lam_{k,j}$ is an eigenvalue of $U_k=D_kC$ corresponding to an eigenvector $\ket{\nu_{k,j}},$ $\ket{\phi_k}=\sum_{x=0}^{n-1} e^{ikx}\ket{x}\in\C^n,$ $k=\frac{2\pi l}{n}, l\in\{0,1,...,n-1\},$ $j\in \{0,1,2\},$ and $D_k=\mbox{diag}(e^{ik},e^{-ik},e^{-ika}).$


\end{proposition}
\noindent{\bf Proof:} First, we consider the image of $\ket{j}\times\ket{\phi_k}$ under $S^{(n,a)}$ for $j=0,1,2.$ For $j=0$ \beano S^{(n,a)}(\ket{0} \otimes \ket{\phi_k}) &=& \sum_{x=0}^{n-1} S_x^{(n,a)} (\ket{0} \otimes \ket{\phi_k}) \\ &=& \ket{0}\otimes \left( \ket{n-1} + \ket{0} e^{ik} + \hdots + \ket{n-3}e^{ik(n-2)} +\ket{n-2}e^{ik(n-1)}\right)\\ &=& \ket{0} \otimes \left(  \ket{n-1}e^{ik+i(n-1)k} + \ket{0} e^{ik} + \hdots + \ket{n-3}e^{ik+ik(n-3)} +\ket{n-2}e^{ik+ik(n-2)} \right)\\ &=& \ket{0}\otimes e^{ik}\ket{\phi_k} =e^{ik} (\ket{0}\otimes \ket{\phi_k}).\eeano  Similarly, $$S^{(n,a)}(\ket{1} \otimes \ket{\phi_k})=e^{-ik} (\ket{1}\otimes \ket{\phi_k}) \,\, \mbox{and} \,\, S^{(n,a)}(\ket{2} \otimes \ket{\phi_k})=e^{-ika} (\ket{2}\otimes \ket{\phi_k}).$$

Let $\ket{\nu_{k,j}}$ be an eigenvector of $U_k$ corresponding to the eigenvalue $\lam_{k,j},$ $j=0,1,2.$  Suppose $C\ket{\nu_{k,j}}=\sum_{j=0}^2 \alpha_j\ket{j}$ for some $\alpha_j\in\C.$ Then \beano U_{(n,a)}\ket{\psi_{k,j}} &=& S^{(n,a)} (C\ket{\nu_{k,j}} \times \ket{\phi_k}) \\ &=& S^{(n,a)}\left(\sum_{j=0}^2 \alpha_j\ket{j} \otimes \ket{\phi_k}\right) \\ &=& \left( \alpha_0 e^{ik}\ket{0} + \alpha_1 e^{-ik} \ket{1}+ \alpha_2 e^{-ika}\ket{2}\right) \otimes \ket{\phi_k}\\ &=& D_kC\ket{\nu_{k,j}} \otimes \ket{\phi_k} \\ &=& \lam_{k,j} \ket{\psi_{k,j}}.\eeano This completes the proof. \hfill{$\square$}

Here we mention that the above proposition is proved in [Lemma 1, \cite{Sadowski2016}] for the case when $C=G,$ the Grover matrix. However, there is a typo in the statement of Lemma 1 about $\ket{\phi_k}.$ Below, we investigate periodicity of lively quantum walks when the coin operator belongs to subsets of $\calPO_R.$ Since periodicity of the walk is decided by the eigenvalues of $U_{(n,a)}$ that are indeed eigenvalues of $U_k,$ we derive eigenvalues of $U_k=D_kC, C\in\calPO_R$, and hence determine the periodicity of the corresponding walk. First, we recall the following results for lazy quantum walks on cycles  \cite{Kajiwara2019}. 

\begin{theorem} \label{thm1}
Consider the lazy quantum walk corresponding to the coin operator $C=[c_{ij}]_{3\times3 }.$ If $c_{11} \not \in \frac{\mathbb{Z}[\zeta_{lcm(n,t)}]}{n}$ or $c_{22} \not \in \frac{\mathbb{Z}[\zeta_{t}]}{n}$  or $c_{33} \not \in \frac{\mathbb{Z}[\zeta_{lcm(n,t)}]}{n}$, then ${U_{(n,0)}}^t \neq I$ for any $t \in \mathbb{N}$.
\end{theorem}
\begin{lemma}\label{lem:kaji}
If $U_{(n,0)}^t=I$ for some positive integer $t$ then $\lam_1(l)+\lam_2(l)+\lam_3(l)\in\Z[\zeta_t]$ where $l\in\{0,1,\hdots,n-1\}$ and $\lam_j(l)$ are eigenvalues of $U_k, k=\frac{2\pi l}{n},$ $j=1,2,3.$
\end{lemma}

A straightforward generalization of the above theorem provides a necessary condition for finite periodicity for lively $(a\neq 0)$ quantum walks on cycles can be stated as follows.  

\begin{lemma}
 Let $C=[c_{ij}]_{3\times3 }$ be a coin operator of order $3$ in a lively quantum walk with liveliness $'a'$ where $n|la $ for $l \in \{0,1,...,n-1\}$. If $c_{11} \not \in \frac{\mathbb{Z}[\zeta_{lcm(n,t)}]}{n}$ or $c_{22} \not \in \frac{\mathbb{Z}[\zeta_{lcm(n,t)}]}{n}$  or $c_{33} \not \in  \frac{\mathbb{Z}[\zeta_{t}]}{n}$, then ${U_{(n,a)}}^t \neq I$ for any $t \in \mathbb{N}$.
\end{lemma}
\noindent{\bf Proof:} The proof follows from the fact that  $e^{ika}=1$ when $n$ divides $la,$ where $k=\frac{2\pi l}{n}$ and $l \in \{0,1,...,n-1\}$ and Theorem \ref{thm1}. \hfill{$\square$}

\subsection{$C\in \calX_R$}

In this section we derive periodicity of the lively quantum walks defined by the walk operator $U_{(n,a)}=S^{(n,a)}(C\otimes I)$  when $C\in\calX_R.$  

\begin{lemma} \label{l2}
Eigenvalues of $U_k=D_kC,$ $C\in\calX_R$ are given by $1$ and $$\lam_k^{\pm} = \frac{x-1}{2}+x\cos{k} \pm \frac{\sqrt{(x+2x\cos{k}+1)(x+2x\cos{k}-3)}}{2}$$ where $-\dfrac{1}{3}\leq x\leq 1,$ and $\frac{n}{\mbox{gcd}(n,a)}$ divides $l\in\{0,1,\hdots,n-1\},$ $k=\frac{2\pi l}{n}.$ Moreover, $\lam_{k}^{\pm}=\lam_{2\pi-k}^{\pm}.$
\end{lemma}
\noindent{\bf Proof:} Note that  $$U_k=diag(e^{ik}, e^{-ik}, e^{-ika})C=\bmatrix{e^{ik}x & e^{ik}y & e^{ik}(1-x-y) \\ e^{-ik}(1-x-y) & e^{-ik}x & e^{-ik}y \\ e^{-ika}y & e^{-ika}(1-x-y) & e^{-ika}x}$$  where $x^2+y^2-x-y+xy=0, -\dfrac{1}{3}\leq x\leq 1.$ Then the characteristic polynomial of $U_k$ is given by $$\chi(\lam) =\lambda^3 -x(e^{-ika}+2\cos{k})\lambda^2+x(e^{-ika}(2\cos{k})+1)\lambda - e^{-ika}.$$ Further, $e^{ika}=1$ since $\frac{n}{gcd(n,a)}$ divides $l$ implies $n$ divides $la.$ Consequently,  $$\chi(\lam) = (\lambda-1)\left(\lambda^2-\lambda(x+2x\cos{k}-1)+1\right).$$ Then the roots of $\chi(\lam)=0$ are $$1,\frac{x}{2}+x\cos{k} \pm \frac{\sqrt{(x+2x\cos{k}+1)(x+2x\cos{k}-3)}}{2}-\frac{1}{2}.$$ 
Hence the desired result follows. \hfill{$\square$}

Now we show that the Grover coin operator is special among all coin operators in $\calX_R\setminus\{I\}$ in terms of its eigenvalues. 

\begin{theorem} \label{thm2}
All the eigenvalues of a matrix $C\in\calX_R$ are real if and only if $C=G=\frac{2}{3}J-I,$ the Grover matrix or $C$ is the identity matrix.  
\end{theorem}
\noindent{\bf Proof:} 
Using Lemma \ref{l2} the eigenvalues of  $U_k$ for  $C\in \mathcal{X}_R$ are $$1,\frac{x}{2}+x\cos{k} \pm \frac{\sqrt{(x+2x\cos{k}+1)(x+2x\cos{k}-3)}}{2}-\frac{1}{2},$$ where $-\frac{1}{3}\leq x\leq 1, k=\frac{2\pi l}{3}$ and  $l=0,1,2.$ Now $k=0$ when $l=0.$ Then $U_0=diag(1,1,1)C$. Thus $U_0=C$, and hence the eigenvalues of $C$ are $$\frac{3x-1}{2}\pm \frac{\sqrt{(3x+1)(3x-3)}}{2}.$$ Obviously $(3x+1)(3x-3)\in\R$ if and only if $x\in\{-\frac{1}{3}, 1\}.$ Thus either $C=I$ or $C=G.$ \hfill{$\square$}

The following theorem describes the periodicity of lively quantum walks when the coin operator belongs to $\calX_R.$  The proof is similar to the proof of [Theorem 2.2, \cite{Kajiwara2019}].

\begin{theorem}\label{periodicity1}
The period of a lively quantum walk on $C_n$ defined by the walk operator $U_{(n,a)}=S^{(n,a)}(C\otimes I), C\in\calX_R=\calX_\theta$ is given by $$ \left\{
  \begin{array}{ll}
    \mbox{lcm}\{c_lp_l : 0< l\leq n-1\} \, \mbox{where} \, \frac{2l}{n}=\frac{m_l}{p_l}, \, \mbox{gcd}(m_l,p_l)=1, \\ 
c_l=1 \, \mbox{if}\,\, m_l \, \mbox{is even and} \, c_l=2, \, \mbox{if}\,\, m_l \, \mbox{is odd}, \,\,\,\,\, \mbox{if} \,\, C=I \\\\
  3, \,\,\,\,\  \mbox{if}\,\, C \in \{P_2, P_3\} \\
\mbox{lcm}\{3, q\}, \,\,\,\,\ \mbox{if} \,\, n=3 \,\, \mbox{and} \,\,\theta=\frac{2m\pi}{q}, m/q\in\Q, \,  \mbox{gcd}(2m,q)=1, \,C\notin\{I,P_2,P_3\} \\
\mbox{lcm}\{3, 2q\}, \,\,\,\,\ \mbox{if} \,\, n=3 \,\, \mbox{and} \,\,\theta=\frac{(2m+1)\pi}{q}, (2m+1)/q\in\Q,  \,  \mbox{gcd}(2m+1,q)=1, \,C\notin\{I,P_2,P_3\} \\
\infty, \,\,\,\, \mbox{otherwise.}
  \end{array}
\right.$$
\end{theorem}


\noindent{\bf Proof:} The characteristic polynomial of $U_k=D_kC$, $C\in \mathcal{X}_R$ is $\chi(\lam)=\lambda^3 -\lambda^2(x+2x\cos{k})+ \lambda(x+2x\cos{k})-1.$ Then setting $x=\frac{2\cos\theta+1}{3}$ and considering the one parameter representation $\calX_\theta$ of $\calX_R$ we proceed as follows. 


First consider the specific values of $\theta$ for which $C$ are Permutation matrices. Setting $\theta =0, \frac{2\pi}{3}$ and $\frac{4\pi}{3}$ the corresponding coin operator $C=I, P_3$ and $P_2$ respectively. If $C\in\{P_2, P_3\}$ the eigenvalues of $U_k,$ $k=\frac{2\pi l}{3},$ $l\in\{0,1,\hdots,n-1\}$ are $1, e^{\frac{2\pi i}{3}}, e^{-\frac{2\pi i}{3}}$ by Lemma \ref{l2}. Obviously $3$ is the period of the corresponding walk operator since $3$ is the smallest positive integer for which $U^3_k=I.$ Similarly, for $C=I$ the eigenvalues of $U_k=D_k$ are $1, e^{\frac{2i\pi l}{n}}$ and $e^{\frac{-2i\pi l}{n}}.$ Then the period of the walk is  the smallest positive integer $T$ for which $2lT/n$ is an even integer. Let $2l/n = m_l/p_l$ where $\mbox{gcd}(m_l,p_l)=1$ for $l=1,\hdots,n-1.$ Then $T=\mbox{lcm}\{c_lp_l : 0< l\leq n-1\}$ where $c_l=1$ if $m_l$ is even and  $c_l=2$ if $m_l$ is odd.   

Next let $\theta\notin \{0,  \frac{2\pi}{3}, \frac{4\pi}{3}\}.$ Then we consider two cases. First assume that $n=3.$ Then $k\in\{0, 2\pi/3, 4\pi/3\}.$ If $l=0,$ that is, $k=0$ then the eigenvalues of $U_k$ are $1, e^{i\theta}, e^{-i\theta}.$ For $l=1,2$ the eigenvalues of $U_k$ are $1, e^{\frac{2\pi i}{3}}, e^{\frac{-2\pi i}{3}}$ which are independent of $\theta,$ and hence $U_k^3=I.$ Otherwise, for $l=0,$ for finite periodicity, $\theta$ has to be of the form $\pm2m/q\pi$ or $\pm (2m+1)/q$ for any nonnegative integer $m.$ Since otherwise if $\theta$ is an irrational number, there can not exist a positive integer $t$ for which $e^{i\theta t}=1.$ Now if $\theta=2m\pi/q$ then the eigenvalues of $U_k$ are $1, e^{\frac{2m\pi i}{q}}$ and $e^{-\frac{2m\pi i}{q}}.$ Hence the period of the corresponding walk is $\mbox{lcm}\{3, q\}.$ On the other hand, if $\theta=(2m+1)\pi/q$ then the eigenvalues of $U_k$ are $1, e^{\frac{(2m+1)\pi i}{q}}$ and $e^{-\frac{(2m+1)\pi i}{q}}.$ Consequently, the period of the corresponding walk is $\mbox{lcm}\{3, 2q\}.$

 Now consider $n \neq 3.$ Then by Lemma \ref{lem:kaji}, if $U_{(n,a)}^t=I$ for some positive integer $t$ then $\sum_{j=1}^3 \lambda_j(l) \in \mathbb{Z}[\zeta_t]$ where $\lam_j(l)$ are eigenvalues of $U_k,$ $k=\frac{2\pi l}{n}$ $j=1,2,3,$ $l\in\{0,1,\hdots n-1\}.$ This yields, $\lambda_2(l)+\lambda_3(l)=(x+xe^{ik}+xe^{-ik}-1) \in \mathbb{A}[\zeta_n]$ and hence $(x+ xe^{ik}+xe^{-ik}-1) \in \mathbb{Z}[\zeta_n].$ Therefore,
 $$\frac{(2\cos{\theta}-2)}{3}+\frac{(2\cos{\theta}+1)}{3}e^{\frac{2\pi l}{n}i}+\frac{(2\cos{\theta}+1)}{3}e^{-\frac{2\pi l}{n}i} \in \mathbb{Z}[\zeta_n]$$ which has to be satisfied for all $l=0,1,2,\hdots, n-1.$

However, for $l=1,$ if this condition is satisfied, we obtain $$\frac{(2\cos{\theta}-2)}{3}+\frac{(2\cos{\theta}+1)}{3}e^{\frac{2\pi }{n}i}+\frac{(2\cos{\theta}+1)}{3}e^{-\frac{2\pi }{n}i} \in \mathbb{Z}[\zeta_n].$$
 Consequently,
 $$ e^{\frac{2\pi}{n}i}\left({(2\cos{\theta}-2)}+{(2\cos{\theta}+1)}e^{\frac{2\pi }{n}i}+{(2\cos{\theta}+1)}e^{-\frac{2\pi }{n}i}\right) \in 3\mathbb{Z}[\zeta_n],$$
which implies
$$\left({(2\cos{\theta}-2)}e^{\frac{2\pi }{n}i}+{(2\cos{\theta}+1)}e^{\frac{4\pi }{n}i}+{(2\cos{\theta}+1)}\right) \in 3\mathbb{Z}[\zeta_n].$$

Further, any $\tau \in \mathbb{Z}[\zeta_n]$ can be written as $\sum_{j=0}^{\phi(n)-1} \zeta_n^i$ where $\phi$ to be the Euler totient function. If $\phi(n)>2$, the coefficients of $e^{\frac{2\pi i\times 0}{n}},e^{\frac{2\pi i\times 1}{n}},e^{\frac{2\pi i\times 2}{n}}$ do not belong to $3\mathbb{Z}$ except for $\cos\theta \in \{1,-1/2\}$ that is for permutation matrices  where $\theta\in\{\pm 2\pi/3, 0\}.$ For $\phi(n)\leq2$, that is, for $n\in\{2,4,6\},$ the coin operator $C$ are not  permutation matrices. As above, the proof follows. Indeed, for $n=2,$
$$(2\cos\theta-2)\zeta_2^1+(2\cos\theta+1)(\zeta_2^2+\zeta_2^0)=(2\cos{\theta}+4)\zeta_2^0 \not \in 3\mathbb{Z}[\zeta_2].$$ Similarly for $n=4$ $$(2\cos\theta-2)\zeta_4^1+(2\cos\theta+1)(\zeta_4^2+\zeta_4^0)=(2\cos{\theta}-2)\zeta_4^1 \not \in 3\mathbb{Z}[\zeta_4],$$ and $$(2\cos\theta-2)\zeta_6^1+(2\cos\theta+1)(\zeta_6^2+\zeta_6^0)=(4\cos{\theta}-1)\zeta_6^1 \not \in 3\mathbb{Z}[\zeta_6]$$ for $n=6.$ Hence the necessary condition for finite periodicity is not satisfied. Thus for $n\neq 3$ the walk has no finite period. This completes the proof. \hfill{$\square$}
   
    
Now we restrict our attention to two specific coin operators in $\calX_R=\calX_\theta.$ Indeed for $\theta=\frac{\pi}{2}$ and $\frac{3\pi}{2},$ the corresponding coin operators are given by $$\Delta_1=\bmatrix{
 \frac{1}{3} & \frac{1+\sqrt{3}}{3} & \frac{1- \sqrt{3}}{3}\\
 \frac{1- \sqrt{3}}{3} & \frac{1}{3} & \frac{1+\sqrt{3}}{3}\\
 \frac{1+ \sqrt{3}}{3} & \frac{1- \sqrt{3}}{3} & \frac{1}{3}} \,\, \mbox{and} \,\, \Delta_2=\bmatrix{
 \frac{1}{3} & \frac{1-\sqrt{3}}{3} & \frac{1+\sqrt{3}}{3}\\
 \frac{1+\sqrt{3}}{3} & \frac{1}{3} & \frac{1-\sqrt{3}}{3}\\
 \frac{1-\sqrt{3}}{3} & \frac{1+\sqrt{3}}{3} & \frac{1}{3}}$$ respectively. Moreover, $\Delta_i^2=G,$ the Grover matrix, $i=1,2.$ Then we have the following corollary.

\begin{corollary}
The period of lively quantum walks on cycles $C_n$ corresponding to the coin operator $\Delta_i, i=1,2$ are given by
 $$\left\{
  \begin{array}{ll}
  12, \,\,\,\, \mbox{if} \,\, n=3 \\
  \infty, \,\,\,\, \mbox{otherwise.} 
  \end{array}
\right. $$
\end{corollary}

\subsection{$C\in\calZ_R$}

 Let $C\in \cal {Z}_R.$ Then the characteristic polynomial of $U_k=SC,$ is given by $$\chi(\lam) =1+\lambda^3 -\lambda^2(e^{-ik}+y-ye^{-ik}+i 2x\sin{k})-\lambda(e^{ik}+y-ye^{ik}-i 2x\sin{k}).$$ Obviously, the product of the eigenvalues is $- 1$. Then we have the following theorem. 

\begin{theorem} \label{periodicity2}
The period of the lively quantum walk on a cycle $C_n$ with walk operator $U_{(n,a)}=S^{(n,a)}(C\otimes I),$ $C\in\calZ_R=\calZ_\theta$ is given by $$ \left\{
  \begin{array}{ll}
    \mbox{lcm}\{2, 2p_l :  0< l\leq n-1\} \, \mbox{where} \, \frac{l}{n}=\frac{m_l}{p_l}, \, \mbox{gcd}(m_l,p_l)=1,  \,\,\,\,\, \mbox{if} \,\,  C \in \{P_6, P_4\} \\\\
2  \,\,\,\,\, \mbox{if} \,\,  C =P_5\\
\mbox{lcm}\{2, 2q\}, \,\,\,\,\ \mbox{if} \,\, n=3, \,\,\theta=2\pi\left(\dfrac{1}{3}+\dfrac{p}{q}\right)\,\mbox{for some nonnegative rational number} \,\, \dfrac{p}{q}, \\
\hfill{\mbox{gcd}(p,q)=1, \, \mbox{and} \,\, C\notin\{P_4,P_5,P_6\}} \\
\infty, \,\,\,\, \mbox{otherwise.}
  \end{array}
\right.$$
\end{theorem}
\noindent{\bf Proof:} The proof is similar to the proof of Theorem \ref{periodicity1}. Indeed, consider the one parameter representation $\calZ_\theta$ of $\calZ_R.$ Then, for $\theta=0,$ the coin operator $C=P_4$. For $\theta=\frac{2\pi}{3}$,  $C=P_5,$ and for $\theta=\frac{4\pi}{3}$, $C=P_6$. Thus when $\theta\in\{0, \frac{2\pi}{3}, \theta=\frac{4\pi}{3}\}$ the coin operators are permutation matrices. If $C=P_5,$ the eigenvalues of $U_k=D_kC$ are $1, \pm 1$ for all $l=0,1,\hdots, n-1.$ Hence the period of the corresponding walk is $2,$ since $U_k^2=I.$ Next, the eigenvalues of $U_k$ for $C=P_6$ are $e^{-ik},e^{\frac{ik}{2}},-e^{\frac{ik}{2}}$ where $k=\frac{2\pi l}{n},$ $l\in\{0,1,\hdots,n-1\}$. In other words, the eigenvalues  are $e^{-\frac{2\pi i l}{n}},e^{\frac{i\pi l}{n}},-e^{\frac{i\pi l}{n}}$. If $l > 0$, let $\frac{l}{n}=\frac{m_l}{p_l}$ where $gcd(m_l,p_l)=1$. Then the following three cases arise. If $m_l, p_l$ are odd, then $U_k^{2p_l}=I,$ hence the period is $2p_l.$ Similarly, the same result is true when one of $m_l, p_l$ is odd and the other one is even. For $l=0,$ that is, $k=0$, the eigenvalues of $U_k$ are $1, \pm 1.$ Hence the period of the corresponding walk is $2.$ Thus the desired result follows for $C=P_6.$ Finally, consider $C=P_4.$ Then the eigenvalues of $U_k$ are $e^{\frac{2\pi i l}{n}},e^{\frac{-i\pi l}{n}},-e^{\frac{-i\pi l}{n}}$. Since $e^{{\frac{-i\pi l}{n}}}=e^{{\frac{i\pi l}{n}}}$, then the arguments of periodicity are similar to that of $P_6$. Hence the desired result follows for $C=P_4.$

Now let $\theta\notin\{0, \frac{2\pi}{3}, \frac{4\pi}{3}\}.$  Note that the characteristic equation of $U_k$ can be written as $$\lambda^3-\lambda^2(e^{-ik}+y-ye^{-ik}+ xe^{ik}-xe^{-ik})-\lambda(e^{ik}+y-ye^{ik}+xe^{-ik}-xe^{ik})+1=0.$$ Then,
for any irrational $\theta$, there can not exists $ t\in \mathbb{N}$ such that $ e^{{i\theta}t} = 1$, and hence $\theta \in \mathbb{Q}$. Consequently, $\frac{\theta}{2} -\frac{\pi}{3} = \frac{p \pi}{q}$ for some nonnegative rational number $p/q$ with $gcd(p,q)=1$. 

Now let $n=3.$ Then the characteristic equation is $1+\lambda^3-\lambda-\lambda^2=0$ when $l=0,$ that is, $k=0.$ Thus  the eigenvalues of $U_k$ are $1, \pm 1$ and hence the period of the corresponding walk is $2$. For $l=1,$ and hence $k=\frac{2\pi}{3}$ the eigenvalues of $U_k$ are  $e^{-i(\theta-\frac{2\pi}{3})},e^{i(\frac{\theta}{2}-\frac{\pi}{3})},-e^{i(\frac{\theta}{2}-\frac{\pi}{3})},$ which are of the form $e^{-\frac{2p\pi i}{q}},e^{i(\frac{p\pi}{q})},-e^{i(\frac{p\pi}{q})}$, considering the one parameter representation $\calZ_\theta$ of $\calZ_R,$ and setting $\frac{\theta}{2}=\frac{\pi}{3}-\frac{p\pi}{q}.$ Then assuming $q, p$ as even or odd the desired result follows, that is, the period is $2q.$ For $l=2,$ that is, $k=\frac{4\pi}{3},$  the eigenvalues are $e^{i(\theta-\frac{2\pi}{3})},e^{-i(\frac{\theta}{2}-\frac{\pi}{3})},-e^{-i(\frac{\theta}{2}+\frac{\pi}{3})}$ which are of the form $e^{\frac{2p\pi i}{q}},e^{-i(\frac{p\pi}{q})},-e^{-i(\frac{p\pi}{q})}$ for some nonnegative rational number $p/q.$ Consequently the desired result follows as above, and hence the period of the corresponding walk is $\mbox{lcm}(2, 2q)$ when $n=3.$

Now, let $n \neq 3.$ If possible let $U_{(n,a)}^t=I$ for some positive integer $t$. Then by Lemma \ref{lem:kaji}, $\sum_{i=1}^3 \lambda_i(l) \in \mathbb{Z}[\zeta_t],$ so that $\sum_{i=1}^3 \lambda_i(l) \in \mathbb{Q}[\zeta_n]$. Then from the characteristic polynomial of $U_k,$ $e^{-ik}+y-ye^{-ik}+ xe^{ik}-xe^{-ik} \in \mathbb{Q}[\zeta_n]$. 
Setting $x=\frac{1+2\cos\theta}{3}$ and the corresponding $y$ from equation (\ref{eqn:z}), we obtain  $$e^{-ik}+\frac{(2\cos{\theta}+1)}{3}(e^{ik}-e^{-ik})+\frac{(1-\cos{\theta}+\sqrt{3} \sin{\theta})}{3}(1-e^{-ik}) \in \mathbb{Q}[\zeta_n]$$ for  $l \in \{0,1,2,...,n-1\}.$ 
Whenever $l=1$, $k=\frac{2\pi}{n}$ and using definition of $\mathbb{Z}[\zeta_n]$ we have $$e^{-ik}\left(1-\frac{(2\cos{\theta}+1)}{3}-\frac{(1-\cos{\theta}+\sqrt{3}\sin{\theta})}{3}\right)+e^{ik}\left(\frac{2\cos{\theta}+1}{3}\right)+\frac{(1-\cos{\theta}+\sqrt{3}\sin{\theta})}{3} \in \mathbb{Z}[\zeta_n],$$ 
which implies  
$$e^{-ik}\left(\frac{1-\cos{\theta}}{3}-\frac{\sqrt{3}\sin{\theta}}{3}\right)+e^{ik}\left(\frac{2\cos{\theta}+1}{3}\right)+\frac{(1-\cos{\theta}+\sqrt{3}\sin{\theta})}{3} \in \mathbb{Z}[\zeta_n].$$ 
Therefore,
$$e^{ik}\left(e^{-ik}(1-\cos{\theta}-\sqrt{3}\sin{\theta})+e^{ik}(1+2\cos{\theta})+ (1-\cos{\theta}+\sqrt{3}\sin{\theta})\right) \in 3\mathbb{Z}[\zeta_n],$$ which further implies
$$(1-\cos{\theta}-\sqrt{3}\sin{\theta})+ e^{ik}(1-\cos{\theta}+\sqrt{3}\sin{\theta})+e^{2ik}(1+2\cos{\theta})\in 3\mathbb{Z}[\zeta_n].$$ 

When Euler's totient function $\phi(n)>2,$ the coefficients of $e^{0},e^{\frac{2\pi i}{n}},e^{\frac{4\pi i}{n}} \in 3\mathbb{Z}[\zeta_n]$ if and only if ${\theta}=0,\pm\frac{2\pi}{3}$ that is, the coin is a permutation matrix.
If $\phi(n)\leq2,$ that is, for $n\in\{2,4,6\},$ we consider the following cases. For non-permutation matrices in $\mathcal{Z}_R$:
\begin{enumerate}
    \item $n=2$: $(1+2\cos{\theta}) \not \in 3\mathbb{Z}[\zeta_2]$
    \item  $n=4$: $(-3 \cos \theta - \sqrt3 \sin \theta)+(1-\cos{\theta}+\sqrt{3}\sin{\theta}) \zeta_4^1 \not \in 3\mathbb{Z}[\zeta_4]$
    \item $n=6$: $(1-\cos{\theta}-\sqrt{3}\sin{\theta})\zeta_6^1+(1+2 \cos \theta) \zeta_6^2 \not \in 3\mathbb{Z}[\zeta_6]$.
\end{enumerate}
The above equations are satisfied only for all permutation matrices in $\mathcal{Z}_R$. However the conditions not satisfied otherwise, when $l=1$. This completes the proof.  \hfill{$\square$}

\subsection{$C\in\calY_R$}
The characteristic polynomial of $U_k=D_kC,$ $C\in\calY_R$ is given by $$\chi(\lam)=\lambda^3+1-\lambda^2(x+2x\cos{k})-\lambda(x+2x\cos{k}).$$ Then the following theorem describes the periodicity of the corresponding quantum walk.
\begin{theorem}\label{periodicity3}
The period of the lively quantum walk on a cycle $C_n$ with walk operator $U_{(n,a)}=S^{(n,a)}(C\otimes I),$ $C\in\calY_R=\calY_\theta$ is given by $$ \left\{
  \begin{array}{ll}
    \mbox{lcm}\{2, 2p_l : 0< l\leq n-1\} \, \mbox{where} \, \frac{2l}{n}=\frac{m_l}{p_l}, \, \mbox{gcd}(m_l,p_l)=1,  \,\,\,\,\, \mbox{if} \,\,  C =-I \\\\ 
6  \,\,\,\,\, \mbox{if} \,\,  C \in\{-I, -P_2, -P_3\}\\
\mbox{lcm}\{6, 2q\}, \,\,\,\,\ \mbox{if} \,\, n=3, \,\,\theta=\frac{m\pi}{q}\,\mbox{for some rational number} \,\, \dfrac{m}{q}, \\ \hfill{\,\,\mbox{gcd}(m,q)=1,\, \mbox{and} \,\, C\notin\{-P_2, -P_3\}} \\
\infty, \,\,\,\, \mbox{otherwise.}
  \end{array}
\right.$$
\end{theorem}
\noindent{\bf Proof:} The proof is similar to the proofs of Theorem \ref{periodicity1} and Theorem \ref{periodicity2}. \hfill{$\square$}

\subsection{$C\in\calW_R$}

The characteristic polynomial of $U_k=D_kC,$ $C\in\calW_R$ is given by $$\chi(\lam) = \lambda^3+\lambda^2(e^{-ik}-y+ye^{-ik}+ -xe^{ik}+xe^{-ik})-\lambda(e^{ik}-y+ye^{ik}-xe^{-ik}+xe^{ik})-1.$$ Then we have the following theorem.
\begin{theorem}\label{periodicity4}
The period of the lively quantum walk on a cycle $C_n$ with walk operator $U_{(n,a)}=S^{(n,a)}(C\otimes I),$ $C\in\calW_R=\calW_\theta$ is given by $$ \left\{
  \begin{array}{ll}
    \mbox{lcm}\{2,2p_l : 0< l\leq n-1\} \, \mbox{where} \, \frac{l}{n}=\frac{m_l}{p_l}, \, \mbox{gcd}(m_l,p_l)=1,  \,\,\,\,\, \mbox{if} \,\,  C \in\{-P_4, -P_6\} \\\\ 
2  \,\,\,\,\, \mbox{if} \,\,  C =-P_5\\
\mbox{lcm}\{2, 2q\}, \,\,\,\,\ \mbox{if} \,\, n=3, \,\,\theta=2\pi\left(\dfrac{m}{q}-\dfrac{1}{6}\right)\,\mbox{for some rational number} \,\, \dfrac{m}{q},\\ 
\hfill{ \mbox{gcd}(m,q)=1,\,\, \mbox{and} \,\, C\notin\{-P_4, -P_5,-P_6\}} \\
\infty, \,\,\,\, \mbox{otherwise.}
  \end{array}
\right.$$
\end{theorem}
\noindent{\bf Proof:} The proof is similar to the proofs of Theorem \ref{periodicity1} and Theorem \ref{periodicity2}. \hfill{$\square$}

\subsection{$C\in\calPO_Q$ }

Now we consider coins which belong to $\calPO_Q=\calX_Q\cup\calY_Q\cup\calZ_Q\cup\calW_Q,$ the group of all rational matrices which are linear sum of permutation matrices of order $3\times 3.$ Obviously, $\mathsf{Grov}\subset\calPO_Q,$  $\calP\subset\calPO_Q$ and the Grover coin $G=\frac{2}{3}J-I\in\mathsf{Grov}.$ In \cite{Kajiwara2019}, it has been shown that period of lively quantum walks on cycles $C_n$ with Grover coin is finite, which is $6,$ if and only if $n=3.$ We have already seen that period is finite when the coin operator $C\in\calP$ or $-C\in\calP,$ see Theorem \ref{periodicity1} - \ref{periodicity4}. We mention that $\calPO_Q \subset \calPO_R$ and hence the periodicity can be calculated for a particular coin $C\in\calPO_Q.$ However, the following result is remarkable since it proves the importance of rational coins which are of Grover-type matrices. We recall Niven's theorem which states that if $\frac{x}{\pi}$ and $\sin{x}$ are both rational, then $\sin x\in \{0,  \pm \frac{1}{2},  \pm 1\} $\cite{Niven}.

\begin{theorem}
Period of lively quantum walks on cycles $C_n$ defined by the walk operator $U_{(n,a)}=S^{(n,a)}(C\otimes I),$ $C\in\calPO_Q$ is finite if and only if $C\, \mbox{or}\, -C\in\calP \cup \mathsf{Grov}.$ In particular, the period of the quantum walk corresponding to the coin operators $C$ or $-C$ in $\mathsf{Grov}$ are given by $$\left\{
  \begin{array}{ll}
    6, \,\,\,\, \mbox{if} \,\, n=3 \,\,\mbox{and},\,\, C \, \mbox{or}\, -C\in\left\{\frac{2}{3}J-P_2, \frac{2}{3}J-P_3\right\}  \\
   12, \,\,\,\, \mbox{if} \,\, n=3 \,\,\mbox{and},\,\, C\, \mbox{or}\, -C=\frac{2}{3}J-P_6\\
  4, \,\,\,\, \mbox{if} \,\, n=3 \,\,\mbox{and},\,\, C\, \mbox{or}\, -C=\frac{2}{3}J-P_5.
  \end{array}
\right. $$
\end{theorem}

\noindent{\bf Proof:} First consider $C\in\calX_Q.$ Then the eigenvalues of $U_k$ that are not $1$ are given as $$\frac{x}{2}+x\cos{k} \pm \frac{\sqrt{(x+2x\cos{k}+1)(x+2x\cos{k}-3)}}{2}-\frac{1}{2}$$ which must be of the form $e^{\pm i\pi \theta}$ and $\cos{\pi\theta}=\frac{x}{2}+x\cos{k}-\frac{1}{2}$, for some $\theta\in\R.$ Finite periodicity of the corresponding walk implies $U_{(n,a)}^t=I$ for some positive integer $t$ which further implies $e^{\pm i \pi t\theta}=1$. Consequently, $\theta$ must be rational. By Niven's Theorem, it can be concluded that when $\theta$ and $\cos{\pi \theta}$ are rational, $\cos{\pi \theta}\in \{0,\pm 1,\pm \frac{1}{2}\}$. Using the description of $\mathcal{X}_{\theta},$ when $k=0$, $x \in$ \{$\pm \frac{1}{3},\frac{2}{3},1,0\}$. Thus the matrices in $\mathcal{X}_R$ are the  permutation matrices or Grover Matrix or the matrices $\frac{2}{3}J-{P_2}$ and $\frac{2}{3}J-{P_3}$.
However, if $x=\frac{1}{3},$ the corresponding matrix $C$ is no longer rational. Hence, only non-permutation matrices in $\mathcal{X}_Q$ that have finite period in quantum lively walk on $C_n$ are $\frac{2}{3}J-I$,$\frac{2}{3}J-P_2$ and $\frac{2}{3}J-P_3$ and $n=3$. Similarly, considering the coin operators belonging to $\mathcal{Y}_R,\mathcal{Z}_R$ and $\mathcal{W}_R$, the desired result follows. The periods follows from Theorem \ref{periodicity1}-\ref{periodicity4}. \hfill{$\square$} \\\\

\noindent{\bf Conclusion.} We have derived period of three state lively quantum walk on cycle when the coin operator of the walk is a real orthogonal matrix which can be expressed as linear sum of permutation matrices. This is achieved by first classifying all orthogonal matrices of order $3\times 3$ that can be expressed as linear sum of permutation matrices. We have also established that the set of such matrices is same as the group of orthogonal permutative matrices. 

We mention that Grover matrix has been used in literature as coin operator to define many quantum mechanical algorithms like quantum walks. This work extends the scope of generalizing all such algorithms by generalizing the Grover matrix into a (real) linear sum of permutation matrices. The limiting distribution of the three state lively quantum walk on cycle and line with generalized Grover coins will be reported in a forthcoming work of the authors.\\\\

\noindent{\bf Acknowledgement.} The authors thank Shantanav Chakraborty for careful reading an earlier version of the manuscript and making several helpful comments. Rohit Sarma Sarkar acknowledges support through junior research fellowship of Ministry of Human Resource and Development (MHRD), India. Amrita Mandal thanks Council for Scientific and Industrial Research (CSIR), India for financial
support in the form of a junior/senior research fellowship.


\end{document}